\documentclass[12pt,letterpaper]{article}
\pdfoutput=1

\usepackage[colorlinks=false,
   linkcolor=red, 
   citecolor=blue,
   filecolor=red,
   urlcolor=red,
   linktoc=page,  
   pdfstartview=FitV,
   bookmarksopen=true]{hyperref}
   
\usepackage[utf8]{inputenc}
\usepackage[left=2cm,top=1cm,right=3cm,nohead]{geometry}
\usepackage[nosort]{cite}
\usepackage{bm}
\usepackage{latexsym,amsfonts,amsmath,amssymb,amsbsy,color,mathrsfs,wasysym,comment,bbold}
\usepackage[dvipsnames]{xcolor}
\usepackage{dsfont}
\usepackage{graphicx,textcomp}
\usepackage{mathtools}

\newcommand{\op}{\hspace{1pt}}
\newcommand{\eq}[1]{\begin{equation}
                     \begin{split} #1 \end{split}
                     \end{equation}}
                     
\numberwithin{equation}{section}

\begin{document}

\thispagestyle{empty}
\begin{flushright}
LMU-ASC 34/17 \\
MPP-2017-112
\end{flushright}
\begin{center}
\end{center}
\vspace{2.2cm}
\begin{center}
{\LARGE \bf{Unwinding strings in semi-flatland}}
 \vskip1.5cm 
Dieter   L\"ust$^{1,2}$, Erik Plauschinn$^{1}$, Valent\'i Vall Camell$^{1,2}$\\
\vskip0.8cm
\textit{$^{1}$ Arnold Sommerfeld Center for Theoretical Physics\\
Theresienstra\ss e 37, 80333 M\"unchen, Germany}\\
\vskip0.4cm
\textit{$^{2}$ Max-Planck-Institut f\"ur Physik\\F\"ohringer Ring 6, 80805
  M\"unchen, Germany}
\vspace{2.5cm}
\end{center}

\begin{abstract}
\noindent
We study the dynamics of strings with non-zero winding number 
around T-duality defects. We deduce that the physics near the core of
such non-geometric objects involves winding modes that are not captured
by the supergravity approximation, and we argue that such corrections
are T-dual to the modes responsible for quantum corrections of
semi-flat elliptic metrics. We furthermore construct a solution of double
field theory that captures part of such near-core physics.
\end{abstract}

\clearpage


\section{Introduction}

It has been known for quite some time that supersymmetric compactifications of string theory on elliptic Calabi-Yau manifolds admit non-geometric modifications, where the elliptic fiber develops monodromies in the U-duality group. In this note we consider perturbative solutions of string theory for which the monodromy is contained in the T-duality group (as part of the U-duality group), and which are generically referred to as T-folds \cite{Hull:2004in}.

A way to construct such spaces is to use an adiabatic fibration of the
higher-dimensional toroidal theory, and to let all toroidal moduli
vary over a base \cite{Greene:1989ya,Hellerman:2002ax,Hellerman:2006tx}. For example,
one can reduce string theory on a $\mathbb T^2$ and then fiber the
complex structure and K\"ahler modulus, that we will denote $\tau$ and
$\rho$ respectively, meromorphically over a $\mathbb{P}^1$ base. If
one only varies $\tau$ the familiar K3 compactifications are recovered, while if also $\rho$ is varied there will be degenerations that generically induce non-geometric monodromies.
In analogy to the geometric case \cite{Greene:1989ya,Strominger:1996it}, we will refer to this situation as a semi-flat approximation.

A semi-flat metric that respects the $U(1)^2$ isometries of the torus fiber can be written as follows (we omit additional transversal directions)
\begin{equation}\label{eq:semiflatmetric}
ds^2  = e^{\varphi} \op d z \op d \bar z + \frac{\rho_2}{\tau_2} \, \bigl|d\xi^1 + \tau\op d\xi^2 \bigr|^2 \, ,
\end{equation}
where $\xi^{1,2}$ are coordinates on the torus and $z$ is a local coordinate on $\mathbb{P}^1$.
Furthermore, we set $\tau = \tau_1 + i \op\tau_2$ and $\rho= \rho_1 +
i\op \rho_2$, and $e^{\varphi}$ is a warp factor. An example which
does not admit a global geometric interpretation is the local solution
around a degeneration of the $\rho$-fibration with monodromy $1/\rho
\rightarrow -1/\rho +1$. In this case, the semi-flat metric
\eqref{eq:semiflatmetric} is usually referred to as a $5_2^2$-brane
\cite{Obers:1998fb,deBoer:2012ma}. More generally, as in the geometric
case, one can have monodromies filling-in the conjugacy classes of
$SL(2,\mathbb{Z})_{\rho}$ \cite{Lust:2015yia}.\footnote{One can also
  consider the T-duality monodromy $(\tau,\rho) \rightarrow
  (\rho,\tau)$, which has been studied in
  \cite{Garcia-Etxebarria:2016ibz} and has been shown to be an essential
  ingredient for the heterotic string away from the stable
  degeneration limit.}

Close to a degeneration of the $\tau$- and $\rho$-fibration this semi-flat approximation breaks down, and the exact local description of the degeneration has to be glued-in.
However, the latter in general breaks (some of) the isometries of the fiber.
As we will review shortly, in the geometric situation we have a good
understanding of such a local description, while in the non-geometric
case the situation is more delicate. In fact, we can rely on a dual
description of the local non-geometric solution, but we will point out
that in all known cases such duality is strictly valid only in the
semi-flat approximation. Given that we lack a conformal field theory description of the degeneration,\footnote{Except particular cases such as asymmetric orbifold points. However, we will be interested in parabolic monodromies which do not admit a CFT description.} and that supergravity is most certainly not valid for such stringy backgrounds, it is important to understand the physics of these degenerations.

By adapting the arguments of \cite{Gregory:1997te} to the torus case,
in this note we argue that such physics is dominated by winding
modes, and that the exotic brane solutions will receive stringy corrections that can be related to the modes correcting the semi-flat ansatz near geometric degenerations.
It has been argued in the literature that modes of this kind can be
captured in a doubled formalism such as double field theory (DFT)
\cite{Hull:2009mi} (for reviews see
\cite{Aldazabal:2013sca,Berman:2013eva,Hohm:2013bwa}). We investigate this by
constructing non-trivial solutions of the DFT
equations of motion that reduce to the ``semi-flat'' exotic brane solution away from the degeneration. However, we did not find any additional justification within the doubled formalism, that this solution captures the correct T-duality of the symmetry-breaking modes.

Finally, in heterotic string theory
we have an additional tool besides T-duality to study the physics of
the $\tau$- and $\rho$-fibrations, as described in
\cite{McOrist:2010jw,Font:2016odl}. If we only consider monodromies
that do not mix the moduli,  the fibrations can be described
algebraically with two elliptic fibrations that encode the solutions
for the varying moduli. In this form, one can use the explicit
heterotic/F-theory duality map \cite{LopesCardoso:1996hq} to construct
the dual K3-fibered Calabi-Yau manifold. Since the moduli are mapped
to geometric moduli of the fiber K3 (which is itself elliptically
fibered), the dual F-theory model is geometric and can be used to
read-off the physics of the non-geometric heterotic
background. However, even in the simple case of a NS5-brane, this map
misses precisely the information about the position of the brane on
the fiber torus, and thus it cannot be used to understand the
near-core physics of the $\rho$-degenerations. It would be 
interesting to understand how these corrections are seen in the
F-theory dual and to compare such dual description with the double
field theory solutions.

\bigskip
This note is organized as follows. In section~\ref{sec_review} we
review the exact metric
for a Kodaira $I_1$ degeneration and for a dual NS5-brane localized on an
elliptic curve, and discuss the dynamics of winding modes and the
T-duality of massive fields.
In section~\ref{sec_t-fold} we extend this analysis to non-geometric solutions, and in section~\ref{sec_dft} we 
investigate a double field theory description. 
Section~\ref{sec_concl} contains a discussion of our results. The
appendix summarizes additional results on domain wall solutions and
non-geometric R-spaces.


\section{Exact metrics and T-duality}
\label{sec_review}

In this section, we review the exact metrics of monopoles 
with compact transverse directions and  study the dynamics of unwinding strings in 
the semi-flat limit of elliptic fibrations. We use the same conventions as in \eqref{eq:semiflatmetric},
for which 
the complex structure and K\"ahler modulus of a two-torus are expressed in terms of the metric and 
Kalb-Ramond field on the $\mathbb T^2$ as
\begin{equation}
\tau=\frac{g_{12}}{g_{22}}+i\,\frac{\sqrt{\det g}}{g_{22}}\, , 
\hspace{50pt}
\rho=B_{12}+i\op\sqrt{\det g} \, .
\end{equation}


\subsection{$I_1$ degeneration}\label{sec:TaubNUT}

In the geometric setting, the simplest semi-flat solution corresponds
to the Kodaira type $I_1$ singularity, which is uniquely determined by the
monodromy acting on the fiber torus when encircling the singularity.  The monodromy is a Dehn twist of fixed
chirality around the shrinking cycle, sending $\tau\rightarrow
\tau+1$. In this case, we know that the exact metric is that of a
Taub-NUT space with one transverse compact direction. The exact metric breaks one of the $U(1)\times U(1)$ isometries of the semi-flat metric, and the modes that break such isometry are the ones that localize the shrinking cycle on the orthogonal cycle of the torus. We thus see that specifying the type of degeneration is enough to capture the symmetries of the exact solution.

The local metric
can be derived by starting from the Euclidean Taub-NUT solution
and compactifying one base direction. To do so, let us consider the background
\begin{equation}\label{eq:ghansatz}
ds^2 = h(\vec x)\, d\vec x^{\op 2} + \frac{1}{h(\vec x)}\,\bigl(d\xi^2 +\omega \bigr)^2 \, ,
\hspace{50pt}
h(\vec x) = 1+\frac{\tilde{R}_2}{2\op |\vec x|} \, ,
\end{equation}
where $\vec x$ denotes coordinates in $\mathbb{R}^3$
and $\xi^2$ denotes the coordinate on the $S^1$-fiber.\footnote{Here and in the following we omit the additional six space-time directions that make the background into a full ten-dimensional solution of
string theory.} The background is regular if $\xi^2$ has a periodicity of $2\pi\tilde{R}_2$, where $\tilde{R}_2$ is the radius of the fiber at infinity. Note furthermore that at the origin $\vec x =0$ of the base, the cycle of the fiber shrinks to zero size.
The one-form $\omega$ is not closed and encodes the non-triviality of the fibration. It is determined up to shifts by exact forms through the relation $d\omega=\star_3dh$, where the latter ensures that the equations of motion with $H=0$ and $e^{\phi}=g_s={\rm const.}$ are satisfied.

The compactification of this background is  achieved by
considering an infinite array of sources on one of the base directions.
The harmonic function $h(\vec x)$  becomes
\begin{equation}\label{eq:compactifiedKK}
h(r,\xi^1) =1+\sum_{n\in\mathbb{Z}}\frac{\tilde{R}_2}{2\op \sqrt{r^2+(\xi^1-2\pi R_1 n)^2}}\,,
\end{equation}
where we split the three-dimensional radial direction into $|\vec{x}|^2=r^2+(\xi^1)^2$. 
The sum in (\ref{eq:compactifiedKK}) does not converge but can be regularized. After Poisson resummation we obtain the Ooguri-Vafa metric\cite{Ooguri:1996me} described by
\begin{equation}\label{eq:oogurivafa}
h(r,\xi^1) = \frac{\tilde{R}_2}{2\pi R_1} \left[ \log(\mu/r) + \sum_{n\neq 0} e^{i \op n\op \xi^1/R_1}\, K_0\left( \frac{|n|\, r}{R_1}\right) \right],
\end{equation}
with $\mu$ a constant that controls the regulator and absorbs also all other possible constants, for instance the first term in \eqref{eq:compactifiedKK}. $K_0$ is the zeroth-order modified Bessel function of the second kind, whose series expansion for large $r$ reads
\begin{equation}\label{expansionK0}
K_0\left( \frac{|n|\, r}{R_1}\right)  = e^{-\frac{|n|r}{R_1}}\sum_{k=0}^\infty\frac{(-1)^k\Gamma(\frac{1}{2}+k)^2}{\sqrt{\pi} k!}\left(\frac{R_1}{2|n|r}\right)^{k+\frac{1}{2}}\,.
\end{equation}
Hence, the leading semi-flat  term in \eqref{eq:oogurivafa} (i.e. the logarithm) is a good approximation of the exact metric far away from the degeneration point up to
 exponentially suppressed terms. In fact, the semi-flat approximation
 of a smooth K3 -- repaired with the Ooguri-Vafa metric at the 24
 $I_1$ points -- gives a metric that is a good approximation of the
 exact Calabi-Yau metric \cite{GrossK3}. The expression
 \eqref{eq:oogurivafa} can also be derived  (in a simple way) 
 field-theoretically
 \cite{Seiberg:1996ns,Becker:2009df} and (in a complicated way) by solving explicitly the
 Riemann-Hilbert problem with wall-crossing technology\cite{Gaiotto:2008cd}.

The semi-flat approximation of the above background  is a flat two-torus fibration pa\-ra\-meterized by the coordinates $(\xi^1,\xi^2)$
over a two-dimensional base $\mathbb R^2$.
For the latter, we introduce polar
coordinates $(r,\theta)$, and the one-form $\omega$ mentioned above can be brought into the form $\omega^{\text{sf}}=f \op d\xi^1$
with $df=\star_2dh$.  Note that in later calculations we take a gauge where
\begin{equation}
\label{eq:semiflat-omega}
\omega^{\text{sf}}=\frac{\tilde{R_2}}{2\pi R_1}\op \theta \op d\xi^1\,. 
\end{equation}
Furthermore, we observe that after
encircling the defect in the base as $\theta\to\theta+2\pi$, the shift $\omega^{\text{sf}}\rightarrow \omega^{\text{sf}} + \frac{\tilde{R}_2}{R_1}\op d\xi^1$
should be compensated by the shift $\xi^2 \rightarrow \xi^2  -\frac{\tilde{R}_2}{R_1} \op\xi^1$
which, as expected, corresponds to the action of a Dehn twist
on the torus cycles.\footnote{We neglect a constant
  shift which is not captured by the action on the homology \cite{Gaiotto:2008cd}.}
The corrections to the semi-flat term in \eqref{eq:oogurivafa}
explicitly break one of the $U(1)$ isometries of the torus
fiber. 
This affects also the one-form \eqref{eq:semiflat-omega}, which is corrected 
(up to gauge transformations) by modified Bessel functions of the second kind
 as 
\begin{equation}
\omega=\omega^{\text{sf}}-\frac{\tilde{R}_2}{\pi R_1} \op r \sum_{k>0} K_1\!\left(\frac{k\, r}{R_1}\right) \,\sin \!\left(\frac{k\, \xi^1}{R_1}\right) d\theta\, .
\end{equation}

The above analysis can be easily extended to a $I_n$ degeneration.
The solution is given by coalescing $n$ Taub-NUT centers, and the Ooguri-Vafa
corrections \eqref{eq:oogurivafa}, which completely smooth out the
semi-flat metric for $n=1$, now replace the semi-flat singularity
with an $A_{n-1}$ singularity, as expected.


\subsubsection*{Monodromy and unwinding strings}

An important point is that even within the semi-flat approximation,
some ``remnant'' of the corrections \eqref{eq:oogurivafa} survives. In
fact, it is useful to look at the action of the monodromy on the
momentum and winding of strings propagating on the torus fiber.
Recall that for a $SL(2,\mathbb{Z})_{\tau}\times SL(2,\mathbb{Z})_{\rho}$ monodromy ($M_{\tau}, M_{\rho})$, acting on the moduli as
\begin{equation}
\tau \rightarrow M_{\tau}[\tau] \equiv \frac{a \op\tau +b}{c\op\tau + d} \, ,
\hspace{60pt} \rho \rightarrow M_{\rho}[\rho]\equiv \frac{\tilde a \op\rho+ \tilde b}{\tilde c \op\rho + \tilde d} \,,
\end{equation}
the corresponding $O(2,2,\mathbb{Z})$ transformation on the combined momentum($\mathbf{n}$)/winding($\mathbf{m}$) vector $(\mathbf{n}, \mathbf{m})$ is given by
\eq{
\label{eq:monodromynm}
\arraycolsep2pt
\begin{array}{lccrcccrc}
\mathbf{n} &\rightarrow & \tilde a  &\arraycolsep3pt\left( \begin{array}{rr} \hphantom{+}a& \hphantom{+}b\\c& d\end{array}\right)  & \mathbf{n} & + &\tilde b &\arraycolsep3pt\left( \begin{array}{rr} -b& a\\-d& \hphantom{+}c\end{array}\right) & \mathbf{m} \,,
\\[16pt]
\mathbf{m} &\rightarrow &\tilde c&\arraycolsep3pt\left( \begin{array}{rr} -c& -d\\a& b\end{array}\right)  & \mathbf{n} & + &\tilde d &
\arraycolsep3pt\left( \begin{array}{rr} d& -c\\-b& a\end{array}\right) &\mathbf{m} \, .
\end{array}
} 
In the simple case of constant $\rho$, that is say $(\tilde a,\tilde b,\tilde c,\tilde d)=(+1,0,0,+1)$, the above transformation reduces to
\begin{equation}
\mathbf{n} \rightarrow M_{\tau}\, \mathbf{n} \, ,
\hspace{60pt}
\mathbf{m} \rightarrow \bigl(M_{\tau}^t\bigr)^{-1}\, \mathbf{m} \, .
\end{equation}
In our case of interest, namely $\tau \to \tau+1$, we have
\begin{equation}
M_{\tau} =\arraycolsep4pt\left( \begin{array}{rr} 1& 1\\0& 1\end{array}\right)  \,,
\end{equation}
giving the transformation 
\begin{equation}\label{eq:mom-wind-KK}
(n_1 ,n_2) \rightarrow (n_1+n_2, n_2)  \, ,\qquad
(m^1, m^2) \rightarrow (m^1, m^2-m^1) \, .
\end{equation}
We see that momentum along the $(1,0)$-cycle ($\xi^1$-direction) is not conserved for a string that moves around a degeneration point at which the $(0,1)$-cycle ($\xi^2$-direction) shrinks. This is in contrast with the translation invariance of the semi-flat metric along both directions of the torus. The exact metric cures this problem by breaking the $U(1)$ isometry along the $(1,0)$-cycle, as we discussed above.

Note however that winding along the $(0,1)$-direction is also not conserved. This is easy to see by taking a string wrapped along the cycle $(1,1)$. Denoting the world-sheet coordinates by
$(\hat\tau,\hat\sigma)$, we consider the trajectory
\begin{equation}
\arraycolsep2pt
\label{eq:I1trajectory}
\begin{array}{lcl@{\hspace{70pt}}lcl}
\xi^1 &=& 2\pi\op R_1 \op \hat\sigma \, , &
\theta &=& 2\pi \op \hat\tau \, ,\\[4pt]
\xi^2 &=& 2\pi \op \tilde{R}_2 \op \hat\sigma \, , &
r &=& r_0 \, ,
\end{array}
\end{equation}
where $(r,\theta)$ are again polar coordinates on $\mathbb{R}^2$ and $(\xi^1,\xi^2)$ are flat coordinates on $ \mathbb T^2$ with periodicity $(\xi^1, \xi^2) = (\xi^1 + 2\pi R_1, \xi^2 + 2\pi \tilde{R}_2)$.
The monodromy around the defect is a Dehn twist, which corresponds to
cutting the torus along $(1,0)$, rotating by $2\pi$ and gluing it
back. The process of unwinding the string along the $(1,0)$-cycle 
corresponds to the patching
$\xi^2_{(2\pi)}=\xi^2_{(0)}-(\tilde{R}_2/R_1)\op\xi^1_{(0)}$, where
$\xi^a_{(0)}$ and $\xi^a_{(2\pi)}$ are the torus coordinates at
$\theta=0$ and $\theta=2\pi$, respectively. With this transformation the trajectory (\ref{eq:I1trajectory})  unwinds the direction $\xi^2$ at $\theta=2\pi$, which  is the semi-flat version
of  the unwinding trajectory in the Taub-NUT space considered in
\cite{Gregory:1997te}. In the latter case, the string can be unwound by
taking it arbitrarily far-away from the core of the monopole because
the $S^1$ circle is non-trivially fibered over  $S^2$ at spatial infinity in the $\mathbb{R}^3$ base. Such a fibration is in fact the Hopf fibration of a three-sphere. The trajectory in this case takes a string wrapping the fiber and a $S^1\subset S^2$ from the north pole to the south pole, where a rotation of the fiber effectively unwinds the string. Our case is a compactified version of this process. Although far away from the degeneration the space is locally  $\mathbb T^2\times S^1$, the global twist gives it the topology of a nilmanifold, which can be seen as a non-trivial fibration of the $(0,1)$-cycle over the remaining torus $\tilde{\mathbb T}^2 = (1,0) \times S^1$. The non-triviality of such fibration gives the unwinding in our case.

There is yet another way to understand equation \eqref{eq:mom-wind-KK}. In the semi-flat
limit we can quantize the string on the $\mathbb T^2$-fiber, and find for the left- and right-moving momenta the expressions 
\eq{
  \label{mom_lr}
  \bigl( p_{L,R}\bigr)_I = \pi_I \pm (G\mp B)_{IJ} L^J \,, 
  \hspace{50pt} I,J=1,2\,,
}
where $\pi_I$ denotes the canonical momentum and $L^I$ is the winding vector. In the present case, these are given by
\eq{
  \pi_I = \binom{n_1/R_1}{n_2/\tilde R_2} \,,
  \hspace{60pt}
  L^I = \binom{R_1\op m^1}{\tilde R_2\op m^2}\,.
}
Furthermore, when encircling the defect as $\theta\to\theta+2\pi$ the coordinates change as
$(\hat\xi^1,\hat\xi^2) = (\xi^1,\xi^2 - \frac{\tilde R_2}{R_1}\op \xi^1)$, as discussed below
equation \eqref{eq:semiflat-omega}. This gives rise to the diffeomorphism
\eq{
  \arraycolsep3pt
  \Omega^I{}_J = \frac{\partial \hat \xi^I}{\partial \xi^J}  = 
  \left( \begin{array}{cc} 1 & 0 \\ -\frac{\tilde R_2}{R_1} & 1 \end{array} \right) .
}
If we require the spectrum to be invariant under $\theta\to\theta+2\pi$, we see that the momenta $p_{L,R}$ 
appearing in the mass formula
have to be invariant. Recalling then that in the present situation $B_{IJ}=0$ and 
$G(\theta+2\pi) = \Omega^{-T} G(\theta)\, \Omega^{-1}$, we find
\eq{
  0&\overset{!}{=} \Delta \bigl( p_{L,R}\bigr)_I
  \\[4pt]
  &= \bigl(\Omega^T\bigr)_I^{\hspace{5pt}J} \bigl( p_{L,R}(\theta+2\pi)\bigr)_J
  - \bigl( p_{L,R}(\theta) \bigr)_I \\[4pt]
  &=\left[ \bigl(\Omega^T\bigr)_I^{\hspace{5pt}J} \pi_J(\theta+2\pi) - \pi_I(\theta) \right] 
  \pm G_{IJ}(\theta) \left[ \bigl( \Omega^{-1} \bigr)^J_{\hspace{5pt}K}\op
  L^K(\theta+2\pi) - L^K(\theta)   \right] ,
}
which leads to the identifications shown in equation \eqref{eq:mom-wind-KK}.


\subsubsection*{Charge inflow}

As in \cite{Gregory:1997te}, the non-conservation of the winding
charge along the $(1,0)$-cycle is compensated by a radial inflow of
charge towards the degeneration point. This arises from a coupling
between the string and a collective coordinate excitation of the
monopole. 

For the Kaluza-Klein monopole this comes from a gauge
transformation of the $B$-field in terms of the unique (up to a
constant) self-dual two-form \cite{Sen:1997zb}
\begin{equation}
\mathcal{B} = \alpha\op d\Lambda \, ,\hspace{50pt} 
\Lambda = \frac{C}{h} \op\bigl(d\xi^2 + \omega\bigr) \, ,
\end{equation}
where $\alpha$ is a parameter that becomes dynamical at the quantum level. The normalization constant $C$ can be fixed by demanding that $\alpha$ has periodicity of $2\pi/\tilde{R}_2$. After compactification, it is possible to derive an exact expression for $\Lambda$ \cite{Grimm:2012rg}. Here, we will only need the semi-flat limit where $\Lambda$ reduces to 
\begin{equation}\label{eq:KK0mode}
\Lambda = \frac{C}{h}\left[d \xi^2 + \frac{\tilde R_2}{2\pi R_1}\theta \op d\xi^1\right]  ,
\end{equation}
which is indeed a self-dual form for the semi-flat $I_1$ degeneration.
Let us now investigate the coupling of the string trajectory with
$\alpha$. 
For this, we embed the semi-flat supergravity configuration into $4+1$ dimensions and study the dynamics of a string moving in this background, described by the action 
\eq{\label{eq:S-coupling}
&S=S_{\text{sugra}}\\
&\hspace{20pt}
-\frac{1}{4\pi}\int d^5x \int d\hat\rho \, d\hat\tau \,\delta(x^a-X^a)\Bigl[\sqrt{\gamma} \,\gamma^{AB} \op\partial_A X^a\partial_BX^b \op G_{ab} +\epsilon^{AB}\op \partial_A X^a\op \partial_B X^b\,\mathcal{B}_{ab}\Bigr] ,
}
where $G$ and $\mathcal{B}$ are the five-dimensional background fields,
$S_{\text{sugra}}$ is the usual NS-NS supergravity action and we have set $\alpha'=1$.
Promoting  $\alpha$ to $\alpha(t)$ and considering the unwinding trajectory \eqref{eq:I1trajectory}, but leaving the motion along the base directions as arbitrary functions of $\hat \tau$, we find that the dynamics of $\alpha(t)$ can be described in terms of the Lagrangian density
\begin{equation}
\mathcal{L}_{\alpha} = \frac12\,\dot \alpha^2 + K \alpha \left[ h^{-1} \op\frac{d\theta}{dt} + (2\pi - \theta)\,\frac{h'}{h^2} \,\frac{dr}{dt}\right] ,
\end{equation} 
where $K$ is a constant. The corresponding equations of motion are solved by
\begin{equation}
\dot\alpha(t) = K \,\frac{\theta- 2\pi}{h} + \alpha_0 \, ,
\end{equation}
with $\alpha_0$  an integration constant. For trajectories with $r={\rm const.}$, we see that after encircling the defect $\dot \alpha$ increases by $2\pi K/h$. We have therefore checked that a string configuration with initial winding charge $m_2=1$ following the unwinding trajectory \eqref{eq:I1trajectory}, couples non-trivially with the background fields via the zero mode. Along this trajectory the string looses its winding charge but this is compensated by an increase of the kinetic energy of the zero mode. From the point of view of the theory reduced on the unwinding cycle, the winding charge is an electric-type charge associated to the gauge field obtained from the reduction of the $B$-field. With the discussed non-trivial coupling the unwinding trajectory generates an inflow of ``winding'' current which is eventually absorbed by the brane configuration \cite{Gregory:1997te}. We will come back to this point below when discussing the T-dual configurations.\\

Finally, generalizing our above discussion, there exist configurations in which an arbitrary $(p,q)$-cycle shrinks when encircling the singularity. A Dehn twist around a $(p,q)$-cycle is represented by the monodromy
\begin{equation}
M_{\tau}^{(p,q)} = \begin{pmatrix}1+ p\op q & p^2\\ -q^2 &1- p\op q \end{pmatrix} ,
\end{equation}
which is the generic element of the parabolic conjugacy class of
$SL(2,\mathbb{Z})$. This corresponds to a compactification of a
Taub-NUT space where the coordinates $(\xi^1, \xi^2)$ have been
rotated by an angle $\phi=\arctan
(q/p)$  compared to the previous example.
To obtain a space which is asymptotically flat one
needs at least 12 mutually non-local $(p,q)$-degenerations, which follows from the minimal Dehn twist decomposition of the identity
\begin{equation}
\bigl(\op M_{\tau}^{(1,0)}\op M_{\tau}^{(0,1)}\op \bigr)^6 = \mathds{1} \, .
\end{equation}
If the number of singular fibers is 24, the base space becomes compact and the
total space is a K3 surface.


\subsection{NS5-branes on $\mathbb{R}^2\times T^2$}\label{sec:localisations}

In the semi-flat limit, a fiber-wise T-duality on the shrinking cycle relates the $I_1$ degeneration to 
a NS5-brane with monodromy $\rho \rightarrow \rho + 1$ \cite{Ooguri:1995wj,Kutasov:1995te} (see also \cite{Kimura:2014bea}). In this case, however, the exact solution breaks all the isometries of the fiber torus.


\subsubsection*{Compactification}

To be more concrete, let us  start from the uncompactified NS5-brane background.
Omitting (as in the last section) the six longitudinal space-time directions, it takes the general form
\begin{equation}
\label{eq:NS50}
\arraycolsep2pt
\begin{array}{lcl}
ds^2&=& \displaystyle h(\vec x)\,d\vec x^2 \,,  \\[4pt]
e^\phi &=& \displaystyle g_s\, h(\vec x) \,, \\[4pt]
H_3 &=& \displaystyle \star_4 \op d h(\vec x) \, , 
\end{array}
\hspace{80pt} h(\vec x) = 1+ \frac{1}{|\vec x|^2} \,,
\end{equation}
where $\vec x\in\mathbb R^4$. 
Next, we compactify two of the transversal directions on a two-torus. To this end, we 
split $\mathbb R^4\to\mathbb R^2\times \mathbb T^2$ and introduce 
polar coordinates $(r,\theta)$ on $\mathbb{R}^2$ and 
coordinates $(\xi^1,\xi^2)$ on $\mathbb T^2$. 
The above solution can then be expressed in the following way
\begin{equation}
\label{eq:NS5}
\arraycolsep2pt
\begin{array}{lcl}
ds^2&=& \displaystyle h(r,\xi^1,\xi^2)\,\Bigl[dr^2 + r^2d\theta^2 + (d\xi^1)^2+(d\xi^2)^2\op\Bigr] \,,  \\[8pt]
e^\phi &=& \displaystyle g_s\, h(r,\xi^1,\xi^2) \,, \\[8pt]
H_3 &=& \displaystyle \star_4 \op d h(r,\xi^1,\xi^2) \, , 
\end{array}
\end{equation}
and the function $h$  can be determined
by considering a rectangular lattice of NS5-branes as
\begin{equation}
\label{ns5_comp}
h(r,\xi^1,\xi^2)=1+\sum_{\vec{n}\in\mathbb{Z}^2}\frac{1}{r^2+(\xi^1-2\pi R_1 n_1)^2+(\xi^2-2\pi R_2 n_2)^2}
\,.
\end{equation} 
This sum is not convergent, but can be regulated with a regulator of the form \cite{Becker:2009df,Grimm:2012rg}
 \begin{equation}
\frac{1}{2\pi R_1R_2}\sum_{n\in\mathbb{Z}^*}\frac{1}{ |n|^2} \,.
 \end{equation}
 By subtracting this term from the original function and performing a Poisson resummation we find
\begin{equation}\label{eq:h-2loc}
h(r,\xi^1,\xi^2)=\frac{1}{2\pi R_1R_2}\left[\log\left(\frac{\mu}{r}\right)+\sum_{\vec{k}\in(\mathbb{Z}^2)^{^*}}K_0\left(\lambda\, r\right)e^{-i\left(\frac{k_1\xi^1}{R_1}+\frac{k_2 \xi^2}{R_2}\right)}\right],
\end{equation}
where $(\mathbb{Z}^2)^{^*}=\mathbb{Z}^2-\{(0,0)\}$ and 
$\lambda=\sqrt{\left(k_1/R_1\right)^2+\left(k_2/R_2\right)^2}$. The
same result can be determined in purely field-theoretic terms
\cite{Diaconescu:1997gu,Becker:2009df} from one-loop corrections to
the gauge coupling of the non-linear sigma model, obtained by reducing
a $\mathcal{N}=2$ theory on the torus.
We emphasize that expression \eqref{eq:h-2loc} makes evident the origin of the symmetry-breaking corrections to the semi-flat metric
\begin{equation}\label{eq:h-2smeared}
h(r) = \frac{1}{2\pi R_1 R_2 }\log\left(\frac{\mu}{r}\right) \, ,
\end{equation}
that is, the terms involving $K_0\left(\lambda\, r\right)$ break the
$U(1)^2$ isometry of the background.
Note also that taking the decompactification limit $r,\xi^1,\xi^2\ll R_1,R_2$ in \eqref{ns5_comp}, one recovers the non-compact harmonic function
shown in \eqref{eq:NS50}, that is
\begin{equation}
h (r,\xi^1,\xi^2) = 1+\frac{1}{r^2+(\xi^1)^2+(\xi^2)^2} \,.
\end{equation}
 If we de-compactify only one of the cycles of the torus, say the one corresponding to $\xi^1$, and define as before $|\vec{x}|^2=r^2+(\xi^1)^2$,
we obtain the familiar result for the H-monopole compactified along one direction
\cite{Gauntlett:1992nn,Gregory:1997te}
\begin{equation}\label{eq:Hmonopole}
h\bigl(|\vec{x}|, \xi^2\bigr) = 1+ \frac{1}{2\op R_2 \op |\vec{x}|}\,\frac{\sinh (|\vec{x}|/R_2)}{\cosh(|\vec{x}|/R_2)-\cos (\xi^2/R_2)} \,.
\end{equation}
 This solution encodes the breaking of the $U(1)$ isometry along the cycle which is dual to the shrinking one in the Taub-NUT space.


\subsubsection*{Monodromies}

As in the previous case, we can understand the breaking of the $U(1)$ isometries from the action of the monodromies \eqref{eq:monodromynm}. Since now $\rho$ is varying, there will be a mixing between momentum and winding states. 
Using the general expression shown in \eqref{eq:monodromynm}, we can deduce the action of $\rho \rightarrow \rho + 1$ on the momentum and winding modes as
\begin{equation}\label{eq:NS5nm}
(n_1, n_2) \rightarrow (n_1+m^2 , n_2-m^1) \, , \hspace{50pt} (m^1, m^2) \rightarrow (m^1,m^2) \, ,
\end{equation}
as expected from T-duality. Momentum can now unwind from the T-dual of the $(0,1)$-cycle, with a trajectory dual to \eqref{eq:I1trajectory}
\begin{equation}\label{eq:NS5trajectory}
\arraycolsep2pt
\begin{array}{lcl@{\hspace{70pt}}lcl}
\xi^1 &=& 2\pi R_1 \op \hat\sigma \, , &
\theta &=& 2\pi \op \hat\tau \, ,\\[4pt]
\xi^2 &=& \frac{2\pi}{ R_2} \op \hat\tau \, ,&
r &= & r_0 \, ,
\end{array}
\end{equation}
where we used that $\tilde R_2 = 1/R_2$. Similarly, for the $(1,0)$-cycle we can write
\begin{equation}
\arraycolsep2pt
\begin{array}{lcl@{\hspace{70pt}}lcl}
\xi^1 &=& \frac{2\pi}{R_1} \op \hat\tau \, , &
\theta &=& 2\pi \op \hat\tau \, ,\\[4pt]
\xi^2 &=& -2\pi R_2 \op \hat\sigma \, ,&
r &= & r_0 \, .
\end{array}
\end{equation}
The canonical momenta that generate translations along the fiber directions are
\begin{equation}
\pi_a = \int d\hat\sigma \left[ g_{ab}\op \partial_{\hat\tau} X^b + B_{ab} \op \partial_{\hat\sigma} X^b\right]  \, .
\end{equation}
In order to compute such quantities we can set $h=1$, effectively
putting the brane in a asymptotically flat background.
For the above trajectories we find, respectively,
\begin{equation}
 \pi_2= \frac{1}{R_2}(2\pi-\theta) \,,\hspace{50pt} \pi_1= \frac{1}{R_1}(2\pi-\theta) \, , 
\end{equation}
which indeed vanish after encircling the defect. Accordingly, the exact metric  breaks translational invariance in both directions.

The non-conservation of momentum shown in \eqref{eq:NS5nm} can also be understood
in a fashion similar to the $I_1$-degeneration discussed in the previous section. 
Quantizing the string on the $\mathbb T^2$-fiber in the semi-flat limit, the left- and right-moving momenta are again given by the general expression \eqref{mom_lr}, where
\eq{
  \pi_I = \binom{n_1/R_1}{n_2/R_2} \,,
  \hspace{60pt}
  L^I = \binom{R_1\op m^1}{R_2\op m^2}\,.
}
When encircling the defect as $\theta\to\theta+2\pi$, the coordinates change as
$(\hat\xi^1,\hat\xi^2) = (\xi^1,\xi^2)$ and hence the diffeomorphism is trivial, 
however, now the $B$-field 
 depends non-trivially on $\theta$. 
Demanding again that the spectrum is invariant, we are led to requiring
\eq{
  0&\overset{!}{=} \Delta \bigl( p_{L,R}\bigr)_I
  \\[4pt]
  &= \bigl( p_{L,R}(\theta+2\pi) \bigr)_I - \bigl( p_{L,R}(\theta) \bigr)_I
  \\[4pt]
  &= \Bigl[ \pi_I(\theta+2\pi) + \bigl(B_{IJ}L^J\bigr)(\theta+2\pi)
  - \pi_I(\theta) - \bigl(B_{IJ}L^J\bigr)(\theta) \Bigr] \pm G_{IJ} \op \Bigl[ 
  L^J(\theta+2\pi ) - L^J(\theta) \Bigr] \,,
  \hspace*{-30pt}
}
which gives $L^I(\theta+2\pi) = L^I(\theta)$ and 
$\pi_I(\theta+2\pi) = \pi_I(\theta)+ \frac{1}{R_1R_2}\left( \begin{smallmatrix} 0 & + 1 \\ - 1 & 0 \end{smallmatrix} \right)_{IJ} L^J(\theta)$. Hence, we find the identifications of 
momentum and winding numbers shown in equation \eqref{eq:NS5nm}.


\subsubsection*{Charge inflow}
As we did for the KK-monopole case, we can compute the coupling of the string with the background collective coordinates. In this case, the zero mode dual to \eqref{eq:KK0mode} is a shift along the toroidal coordinate, $\xi^2\rightarrow\xi^2+\alpha$.  In analogy to the above situation, we embed the semi-flat configuration into $4+1$ dimensions and study the dynamics of a string moving in this background using the action \eqref{eq:S-coupling}. Promoting $\alpha$ to $\alpha (t)$ and considering the trajectory \eqref{eq:NS5trajectory} but letting the motion along the angular coordinate on the base be an arbitrary function $\theta(\hat{\tau})$,  the dynamics of the zero modes is described by the effective Lagrangian density
\begin{equation}
\mathcal{L}_\alpha=\frac{1}{2}\dot{\alpha}^2+ \tilde{K}\dot{\alpha}\left(4\pi h -\theta\right),
\end{equation}
where $\tilde{K}$ is a constant. The corresponding equations of motion are solved by
\begin{equation}
\alpha=\tilde{K}\int^t dt \,\theta.
\end{equation}
As for the KK-monopole, after going around the defect $\dot{\alpha}$
increases by $2\pi\tilde{K}$. In this case, the non-conserved charge
along the trajectory is momentum, which couples to the background
fields via the zero mode associated to the position of the brane along
the fiber. Again, one can also perform an analysis from the point of view of the dimensionally
reduced theory. In this case, momentum charge is associated to the KK gauge field coming from the reduction of the metric. The trajectory \eqref{eq:NS5trajectory} will then produce an equivalent current inflow that is absorbed by the background via the discussed mechanism.


\subsubsection*{T-duality of exact metrics}

It is interesting to ask what happens to the T-duality transformation between the
NS5-brane and the $I_1$ degeneration, once corrections to the semi-flat
approximation are taken into account and thus the Buscher rules cannot be applied.

Corrections to the $U(1)$ isometry of the $(1,0)$-cycle have a
physical interpretation related to the non-conservation of
momentum along that cycle. For the $I_1$ degeneration the corrections are
captured by the Ooguri-Vafa metric related to \eqref{eq:oogurivafa} and have the
form of a sum of non-perturbative terms 
\begin{equation}\label{I1corrections}
\mathcal{C}_n \sim e^{-\frac{|n| r}{R_1} }\: e^{-i\op
  \frac{n\op \xi^1}{R_1}} \,,
\end{equation}
where each of these contains also the perturbative sum
\eqref{expansionK0}. The NS5-brane metric related to \eqref{eq:h-2loc} on the other hand contains a double-sum of terms 
\begin{equation}
  \tilde{\mathcal{C}}_{n_1,n_2} \sim 
  e^{-\lambda \op r }\: e^{-i\op  \frac{n_1 \xi^1}{R_1}}\:
  e^{-i\op  \frac{n_2\xi^2}{R_2}} \hspace{70pt} \mbox{with}\quad \lambda = \sqrt{(n_1/R_1)^2
  + (n_2/R_2)^2} \,.
\end{equation}
The corrections depending on $\xi^2$ break the
isometry along the $(0,1)$-cycle, along which we dualize to arrive at
the $I_1$ degeneration. The problem of dualizing these higher Fourier
modes is in fact similar to the problem considered in
\cite{Gregory:1997te}, where it has been suggested (and to some extend
checked in \cite{Harvey:2005ab,Jensen:2011jna}) that the modes in
$\tilde{\mathcal{C}}_{n_1,n_2}$ map to stringy modes of the Taub-NUT
space.
In our case we see that $\tilde{\mathcal C}_{n,0} = \mathcal C_n$ -- including numerical factors --
and it is plausible  to conjecture that T-duality of the
full NS5-background sends each mode $\tilde{\mathcal{C}}_{n,m}$ for $m\neq0$
to a mixed momentum-winding mode $(n,m)$ on the Taub-NUT
side. Note that this is a very specific rule for the massive modes,
and it might be valid only in the regime where the semi-flat
approximation is broken only mildly. 

A similar conclusion is found by considering elements of the T-duality
group that are merely changes of basis, belonging to the geometric
$SL(2,\mathbb{Z})_{\tau}$ subgroup. An example is the rotation that
sends $\tau \rightarrow -1/\tau$, exchanging a $(1,0)$ $I_1$
degeneration with a $(0,1)$ one. The T-duality of exact metrics can
now be derived by noticing that the two configurations are obtained by a
compactification of Taub-NUT spaces along two orthogonal
directions, and  they are therefore related by a $\pi/2$ rotation of the toroidal coordinates. This results in a specific map for the massive modes
\eqref{I1corrections} that sends $\mathcal{C}_n \rightarrow e^{-|n| r/R_2' }\: e^{-i\op
  n\op \xi^2/R_2'}$, with $R_2'=R_1$ and $R_1'=R_2$.\\


\section{T-folds}
\label{sec_t-fold}

Starting from the semi-flat metric of a compactified $I_1$ degeneration -- the
Kaluza-Klein monopole smeared on a $S^1$ discussed in section~\ref{sec:TaubNUT}  
-- we want to perform a T-duality along the circle parametrized by $\xi^1$. At first this seems problematic because the monodromy around
the $I_1$ degeneration acts non-trivially on this $S^1$ and the
corresponding Killing vector $\partial/\partial \xi^1$ is not globally
defined.\footnote{See for example \cite{Belov:2007qj} for a clear
  discussion of these issues.} 
However, the semi-flat NS5-solution has two $U(1)$ isometries and we
can perform a collective T-duality transformation \cite{Plauschinn:2014nha}
or a duality rotation in $O(2,2,\mathbb{Z})$ that corresponds to a fiberwise transformation
$(\rho \rightarrow -1/\rho, \tau\rightarrow-1/\tau)$. An independent
argument showing the appearance of such solutions uses 
heterotic/F-theory duality \cite{McOrist:2010jw,Lust:2015yia}.
The result is the line element
\eq{
\label{eq:Qbrane}
\arraycolsep2pt
\begin{array}{lcl}
ds^2&=&\displaystyle  h(r)\,\bigl(dr^2+r^2d\theta^2\bigr)+\frac{4\pi^2 h(r)}{4\pi^2
      h(r)^2+\tilde{R}_1^2\op\tilde{R}_2^2\op\theta^2}\, \Bigr[
      (d\xi^1)^2+(d\xi^2)^2\Bigr]\, ,\\[10pt]
B&= &\displaystyle -\frac{2\pi\tilde{R}_1\tilde{R}_2\theta}{4\pi^2 h(r)^2+\tilde{R}_1^2\tilde{R}_2^2\theta^2}\, d\xi^1\wedge d\xi^2\,,\\[10pt]
e^{2\phi}&= &\displaystyle \frac{4\pi^2h(r)}{4\pi^2 h(r)^2+\tilde{R}_1^2\tilde{R}_2^2\theta^2}\,,
\end{array}
}
where $\tilde{R}_a=1/R_a$, $(r,\theta)$ parametrize $\mathbb R^2$, and
$h(r)$ is the semi-flat harmonic function \eqref{eq:h-2smeared}. This
solution -- usually denoted as $5^2_2$-brane \cite{deBoer:2012ma} or
called a Q-brane \cite{Hassler:2013wsa} --  clearly induces a
monodromy $-1/\rho \rightarrow -1/\rho +1$, corresponding
to $\beta$-transformations in
$O(2,2,\mathbb{Z})$. It is therefore a globally non-geometric background.

Note that the metric in \eqref{eq:Qbrane} is translational invariant
along both the fiber directions. 
Correspondingly, from \eqref{eq:monodromynm} we deduce 
the non-geometric monodromy action on the momentum and winding states
as
\begin{equation}\label{eq:nmQbrane}
(n_1 , n_2 ) \rightarrow (n_1, n_2) \, ,\hspace{50pt} (m^1, m^2)\rightarrow (m^1 + n_2, m^2-n_1 ) \,.
\end{equation}
This can also be obtained from the NS5-monodromy by the action of the transformation $\rho \rightarrow -1/\rho$ and $\tau\rightarrow-1/\tau$, which interchanges $n_a\leftrightarrow m^a$ for all $a$. 
This suggests that the $U(1)^2$ isometries of the metric \eqref{eq:Qbrane} will not receive quantum corrections, as the momenta are conserved on both the torus directions. However, in analogy with the duality between the Taub-NUT space and the NS5-brane, there exist now trajectories along which strings initially wrapped along the $(1,0)$- and $(0,1)$-cycle of the torus unwind. For example, for the trajectory
\eq{
\arraycolsep2pt
\label{Qtrajectory}
\begin{array}{lcr@{\hspace{70pt}}lcl}
\xi^1 &=& 2\pi \op \tilde R_1 \op\hat\sigma \, ,&
\theta &=& 2\pi \op \hat\tau \, ,\\[4pt]
\xi^2 &=& -\frac{2\pi}{ \op \tilde R_2} \op\hat\tau \, ,&
r &=& r_0 \, ,
\end{array}
}
a string with winding along the $(1,0)$-cycle and momentum along the $(0,1)$-cycle will unwind after encircling the defect.
Note that the monodromy action is similar to the NS5-monodromy
\eqref{eq:NS5nm}, up to an interchange of momenta and windings.

Let us again derive the change in momentum and winding numbers using the invariance of
the left- and right-moving momenta \eqref{mom_lr} when encircling the defect. 
More concretely, we quantize the string on the $\mathbb T^2$-fiber in the semi-flat approximation, 
and find for the canonical momentum and winding vector the expressions
\eq{
  \pi_I = \binom{n_1/\tilde R_1}{n_2/\tilde R_2} \,,
  \hspace{60pt}
  L^I = \binom{\tilde R_1\op m^1}{\tilde R_2\op m^2}\,.
}
Under $\theta\to\theta+2\pi$, the background can be made globally-defined by identifying 
$\mathbb T^2(\theta+2\pi)$ and $\mathbb T^2(\theta)$ using an $O(2,2)$ transformation. 
In particular, we have
\eq{
  \bigl(G\mp B\bigr)(\theta+2\pi) = \mathcal O^{-1}_{\beta} \left[ \bigl(G\mp B\bigr)(\theta) \right]
  \,,
}
where $\mathcal O_{\beta}$ is a so-called $\beta$-transformation. On the combined 
momentum-winding vector $(L^I,\pi_I)^T$ this transformation acts by matrix multiplication as
\eq{
  \renewcommand{\arraystretch}{1.2}
  \mathcal O_{\beta} = \left(
  \begin{array}{c|c} \mathds 1 & 
  \begin{array}{@{}c@{}c@{}} \scriptstyle 0 & \scriptstyle -\tilde R_1 \tilde R_2 \\[-5pt] 
  \scriptstyle +\tilde R_1 \tilde R_2 & \scriptstyle 0 \end{array} 
  \\
  \hline
  0 & \mathds 1
  \end{array}
  \right).
}
We now  demand that the spectrum does not change when encircling the defect, 
which means that the left- and right-moving momenta have to be invariant under
$\theta\to\theta+2\pi$. 
We then compute
\eq{
  0&\overset{!}{=} \Delta \bigl( p_{L,R}\bigr)_I
  \\[4pt]
  &= \mathcal O_{\beta} \bigl( p_{L,R}(\theta+2\pi)\bigr)_I
  - \bigl( p_{L,R}(\theta) \bigr)_I \\[4pt]
  &= \mathcal O_{\beta} \Bigl( \pi_I(\theta+2\pi) \pm  \mathcal O^{-1}_{\beta} \left[ \bigl(G\mp B\bigr)(\theta) \right]_{IJ} L^J(\theta+2\pi) \Bigr)
  - \Bigl( \pi_I(\theta) \pm \left[ \bigl(G\mp B\bigr)(\theta) \right]_{IJ} L^J(\theta)  \Bigr)\,,
  \hspace*{-30pt}
}
leading to the relation
\eq{
  0\overset{!}{=} \mathcal O_{\beta} \cdot \binom{L^I(\theta+2\pi)}{\pi_I(\theta+2\pi)}
  - \binom{L^I(\theta)}{\pi_I(\theta)} \,,
}
which is solved by \eqref{eq:nmQbrane}.

Additionally, the non-conservation of the winding charge should be compensated by an inflow current, and we expect the winding modes to couple to two dyonic coordinates arising from the flux. It is hard to make this concrete because of the non-geometric nature of the local metric \eqref{eq:Qbrane}, but we can make the following argument based on T-duality. 
Let us start from the solution of NS5-branes smeared on the $\mathbb T^2$, and consider the coordinate shifts $\xi^1\rightarrow \xi^1+f_1(r,\theta)$ and $\xi^2\rightarrow \xi^2+f_2(r,\theta)$.\footnote{These
transformations are coordinate transformations, which result in another supergravity solution. 
The zero-mode is a particular case thereof.}
If one applies T-duality along the $\xi^2$-direction to the transformed solution, we see that $f_1$ remains as a coordinate shift of the Taub-NUT solution, while $f_2$ is mapped to a gauge transformation of the $B$-field. This is consistent with the analysis of the monodromy action in section \ref{sec:TaubNUT}. On the other hand, if we start from the NS5-brane configuration and perform two T-dualities, both transformations become gauge transformations of the $B$-field and the metric is not affected. This suggests that the Q-brane has two dyonic zero modes, as expected from T-duality.


\subsubsection*{Beyond semi-flat approximation}

As for the duality between A-type singularities and NS5-branes, we
should ask what is the transformation of the modes \eqref{eq:h-2loc}
that localize the NS5-branes on the fiber torus under the conjectured
T-duality that leads to the solution \eqref{eq:Qbrane}. The answer is
roughly a T-dual version of the transformation between a $(1,0)$ and a
$(0,1)$ type $I_1$ degeneration. A naive guess is that the 
NS5 Fourier modes are mapped to\,\footnote{For the T-duality transformation between the $I_1$ degeneration and the Q-brane this has been checked in \cite{Kimura:2013fda,Kimura:2013zva}.}
\begin{equation}\label{exoticcorrections}
  \tilde{\mathcal{C}}_{n_1,n_2} \sim 
  e^{-\tilde\lambda \op r }\: e^{-i\op  n_1 \tilde\xi_1 \tilde R_1}\:
  e^{-i\op  n_2\tilde \xi_2 \tilde R_2} \hspace{70pt} \mbox{with}\quad \tilde\lambda = \sqrt{(n_1\tilde R_1)^2
  + (n_2\tilde R_2)^2} \, ,
\end{equation}
where we define $\tilde R_i=\frac{1}{R_i}$. As in \cite{Gregory:1997te}, the modes $\tilde \xi_i$
should be identified with dyonic degrees of freedom of the
non-geometric solution, as follows from a particular effective action
describing the type of couplings between winding
and dyonic modes described above. The
rationale for such transformations is that both the geometrical
coordinates $\xi^i = \xi^i_L +\xi^i_R$ and the dual ones $\tilde \xi_i =
\xi^i_L - \xi^i_R$ play a non-trivial role. The semi-flat solution
for a NS5-brane is written in terms of a trivial fibration of the
$(\xi^1,\xi^2)$ fiber coordinates, and the
excited Kaluza-Klein momentum states break both the U(1) symmetries
associated to shifts in such coordinates. The dual stringy coordinates
are instead exact. Note that from the previous discussion it seems
that such stringy coordinates are associated with a non-geometric
fibration structure, so that the present situation is substantially
more complicate than the usual duality between $H$-monopoles and
Taub-NUT spaces. In this latter situation, the stringy coordinate is
associated with a topologically non-trivial circle fibration, which is
traded by T-duality with a $B$-field in the dual, trivially fibered
solution. In fact there is a well-known geometrical construction that unifies
both fibrations \cite{Bouwknegt:2003vb}. Starting from an
oriented $S^1$ bundle over a compact connected manifold $M$: $S^1 \rightarrow E \xrightarrow{\pi} M$, one
constructs the correspondence space $C=E \times_M \hat E$, where $\hat
E$ is the T-dual fibration. $C$ is both a circle bundle over $E$ and a
circle bundle over $\hat E$, and if $\hat E$ is a trivial fibration,
as in the $H$-monopole case, we have that $C= E \times S^1$. For the
present case of elliptic fibrations, this geometric construction
cannot be easily generalized \cite{Belov:2007qj}, in line with the above
discussion. The breaking of both $U(1)^2$ isometries of the NS5-background poses in fact additional challenges for a geometric
description in a extended space, as we will discuss in the next section.


\section{Description in extended space}
\label{sec_dft}

In the previous sections we have seen evidence for a
``generalized T-duality'' acting on higher Fourier modes of the string
fields. We now want to discuss to what extend this physics can be captured by
a T-duality covariant formalism such as the doubled formalism of \cite{Hull:2004in}.


\subsection{Doubled torus fibrations}

In the semi-flat limit, the defect solutions are fully characterized by
associating at each base point a string state $\Psi = \sum \Psi_{n,m}|\textbf{n},\textbf{m}\rangle$ together
with a monodromy, where the latter is an 
$O(2,2,\mathbb{Z})$ transformation acting on the momentum and winding numbers.
We can then Fourier transform the basis $|\textbf{n},\textbf{m}\rangle$ to
position space as
\begin{equation}
|\bm{\xi}, \tilde{\bm{\xi}}\rangle = \sum_{n_i,m_i} e^{i n_1  \xi_1/R_1 +i n_2
  \xi_2/R_2 }e^{i m_1 \tilde \xi^1 \tilde R_1 +i m_2 \tilde \xi^2 \tilde R_2}|\textbf{n},\textbf{m}\rangle
  \,,
\end{equation}
where we  introduced coordinates $(\tilde \xi^1, \tilde \xi^2)$ conjugate to the winding numbers.
These additional coordinates can be thought of as defining an extended compact
space, a four-torus $\mathbb T^4$,
which leads to the doubled formalism of \cite{Hull:2004in}. In our case,
the parabolic monodromies act as generalized Dehn twist on such a
four-torus, and define a (non-principal) fibration over the 
two-dimensional base. For the local model of a Kodaira $I_1$ degeneration
the monodromy \eqref{eq:mom-wind-KK} obviously defines the
$SO(2,2,\mathbb{Z})\subset SL(4,\mathbb{Z})$ monodromy
\begin{equation}\label{monodromyAtau}
A_{\tau} =\begin{pmatrix} M^{(1,0)} & 0\\ 0 & M^{(0,1)}\end{pmatrix}
= \begin{pmatrix} 1& 1&0&0\\ 0&1&0&0\\0&0&1&0\\ 0&0&-1&1\end{pmatrix} \,.
\end{equation}
All the monodromies for the $(p,q)_{\tau}$- and $(p,q)_{\rho}$-defects
can be obtained by a simple change of basis for the $\mathbb T^4$, which
corresponds to different embeddings of two-tori $\mathbb T^2\times \tilde
{\mathbb T}^2$. The $(0,1)$ degeneration, that we will denote by $B_{\tau}$, is
obtained by exchanging $M_{\tau}^{(1,0)}\leftrightarrow M_{\tau}^{(0,1)}$. By exchanging $(\xi_2 \leftrightarrow \tilde
\xi^2)$ or $(\xi_1\leftrightarrow
\tilde\xi^1)$ we obtain respectively the NS5- and Q-brane monodromies
\begin{equation}\label{eq:doubledmonodromy}
A_{\rho}=\begin{pmatrix} 1& 0&0&1\\ 0&1&-1&0\\0&0&1&0\\ 0&0&0&1\end{pmatrix} \,,
\hspace{50pt}
B_{\rho}=\begin{pmatrix} 1& 0&0&0\\ 0&1&0&0\\0&1&1&0\\
  -1&0&0&1\end{pmatrix} \, ,
\end{equation}
corresponding to a B-shift and $\beta$-transform as
expected. Restricting these $\mathbb T^4$ fibrations over the boundary of a
disk around the degeneration, $S^1= \partial D^2$, one recovers the
fibrations of \cite{Hull:2009sg}. In our case these fibrations extend
over the punctured base and we need to ask if there exists a
degeneration of the four-torus giving rise to such monodromies, and if
the local models can be glued together to form a global space. 


\subsubsection*{Singular fibers}

We expect the type of singular fiber in the $\mathbb T^4$ fibration to be
determined by the conjugacy class of the monodromy around the boundary
of a small disk encircling the degeneration. In the doubled
description all the parabolic $\tau$- and $\rho$-monodromies are
related by the change of basis described above, and so all singular fibers
should have the same topology. 
If we assume  that
monodromies of the type \eqref{eq:doubledmonodromy} arise as a
Picard-Lefschetz type monodromy around a singular fiber, where two of
the cycles of the $\mathbb T^4$ are pinched, we obtain a topology of type $I_1
\times I_1$. 
Higher Fourier modes will correct the semi-flat approximation by localizing the
shrinking cycles in a similar way as for the $I_1$ degenerations. The
change of $SL(4,\mathbb{Z})$ duality frames gives then a precise
generalized T-duality between higher Fourier modes, of the kind
discussed in the previous sections. It would be interesting to
study in more details this quantum corrected metric for the $I_1\times I_1$
degeneration.


\subsubsection*{Global construction}

The global issues appear quite subtly, and we would like to make the 
following observation. 
A global, supersymmetric, non-geometric
model can be obtained by pairing 12 non-local $\tau$-degenerations
with 12 non-local $\rho$-degenerations \cite{Hellerman:2002ax}. In the
doubled space this is described by the factorization of the
identity in terms of the $(A,B)$ twists:
\begin{equation}
(A_{\tau} B_{\tau})^6 (A_{\rho}B_{\rho})^6 = \mathds 1 \, .
\end{equation}
However, as a $\mathbb T^4$ fibration (including monodromies in the
full mapping class group $SL(4,\mathbb{Z})$), each degeneration can be
seen as a collision of two elementary degenerations in which one cycle
shrinks. Locally, these correspond to a singular fiber of type
$I_1\times \mathbb T^2$. We see then that the global doubled fibration is
specified by 48 elementary degenerations, which appears to be incompatible with
a holomorphic fibration of the $\mathbb T^4$ moduli. 

This can be seen already
in the geometric setting. Let us consider a smooth K3 surface,
described by 24 mutually non-local $I_1$
degenerations, corresponding to the monodromy decomposition $(M_{\tau}^{(1,0)}
M_{\tau}^{(0,1)})^{12} = \mathds 1$. The doubled torus fibration will then be
described by the decomposition $(A_{\tau}
B_{\tau})^{12} = \mathds 1$. Now, there exists a global polarization that
identifies the physical fiber with the $(\xi_1,\xi_2)$ directions, and
the fibration reduces to a $\mathbb T^2 \times \tilde{\mathbb T}^2$ fibration over
$\mathbb{P}^1$. If we try to fiber the two complex structure moduli
$\tau$ and $\tilde \tau$ of the two tori we can write the metric \cite{Candelas:2014jma,Candelas:2014kma}
\begin{equation}\label{eq:doubletorusfibration}
ds^2  = e^{\varphi}  \op\tau_2\op\tilde\tau_2\op d z  d \bar z + \frac{1}{\tau_2} \, \bigl|d\xi^1
+ \tau \op d\xi^2 \bigr|^2 +\frac{1}{\tilde \tau_2} \, \bigl| d\tilde \xi_1 +
\tilde \tau \op d\tilde \xi_2 \bigr|^2 \, .
\end{equation}
Each $I_1$ degeneration of $\tau$ or $\tilde \tau$ would give the
same deficit angle as the physical $I_1$ singularity we started with,
and a compact model seems to require a total of 24 degenerations,
precisely half of the degenerations required to build the 24
$I_1\times I_1$ degenerations of the doubled model.
We leave this issue for future investigation.


\subsection{Double field theory and generalized duality}\label{sec:DFT-Tduality}

We now discuss solutions of double field theory \cite{Hull:2009mi}
that capture part of the physics discussed in the previous
sections. 
From a slightly different perspective, DFT (and EFT) configurations describing the NS5-brane and its dual backgrounds have been studied also in \cite{Berman:2014jsa,Berman:2014hna,Bakhmatov:2016kfn}.
We denote the coordinates on the ``doubled''  manifold as
$X^N=(x^\mu,\tilde{x}_\mu)$, where the capital indices $N$ are raised and lowered by the $O(d,d)$ metric
\begin{equation}
\eta=\left(\begin{array}{cc}
0&\mathds 1\\
\mathds 1&0
\end{array}\right) .
\end{equation} 
The metric $g$ of the manifold $M$ and the anti-symmetric tensor field $B$ 
are reorganized into the $O(d,d)$ matrix 
\begin{equation}\label{eq:gen-metric}
\mathcal{H}=\left(\begin{array}{cc}
g-B\op g^{-1}B & B\op g^{-1}\\
-g^{-1}\op B & g^{-1}
\end{array}\right),
\end{equation}
also called a generalized metric, and  the dilaton $\phi$ is expressed using the generalized dilaton $\tilde \phi$ defined by
$e^{-2\tilde \phi}=\sqrt{g}\,e^{-2\phi}$. 
In order for the algebra of infinitesimal diffeomorphisms to close, one has to impose 
the so-called strong constraint
\begin{equation}
\eta^{MN}\partial_M\partial_N =0 \,,
\end{equation}
which implies that the fields only depend on half of the generalized
coordinates. Taking this field content, one can write down a manifestly $O(d,d)$-covariant theory \cite{Hohm:2010pp} 
\begin{equation}
\label{eq:DFT-action}
S\sim \int d^NXe^{-2\tilde{\phi}}\,\mathcal{R}(\mathcal{H},\tilde{\phi})\, ,
\end{equation}
where $\mathcal{R}(\mathcal{H},\tilde{\phi})$ is the generalized
curvature scalar (see for instance equation (4.24) in \cite{Hohm:2010pp}).  Solving the strong constraint by demanding no winding-coordinate dependence of the fields, the NS-NS supergravity action is recovered.

Turning to the symmetry transformations, under a global $O(d,d)$-transformations of the form
\begin{equation}\label{eq:general-O(dd)}
\mathsf h=\left(\begin{array}{cc}
A&B\\
C&D
\end{array}\right)\in O(d,d) \,,
\end{equation}
the generalized metric $\mathcal{H}$ transforms as
$\mathcal{H}'(X')=\mathsf h\, \mathcal{H}(X)\, \mathsf h^t$, where $X'=\mathsf h^{-t} X$. 
In terms of the combined background field $E=g+B$, the $O(d,d)$ transformation $\mathsf h$ acts as
\begin{equation}\label{eq:E-transformation}
E'=\mathsf h[E]=(AE+B)(CE+D)^{-1}\,.
\end{equation}
This is the extension of the Buscher rules to arbitrary elements of the
duality group \cite{Giveon:1991jj}. $\Theta$~shifts and
$GL(d)$ transformations are exact, while other elements generically
receive corrections from the path-integral measure. An example of the
latter situation is a T-duality along $x^a$ direction given by the factorized duality
\begin{equation}\label{Erule}
\mathsf h = \left(\begin{array}{cc}
\mathds{1}-e_a&e_a\\
e_a&\mathds{1}-e_a
\end{array}\right),
\end{equation}
where $\mathds{1}$ is the $d\times d$ identity matrix and $e_a$ is a
$d\times d$ matrix with all entries equal zero except for the a'th
diagonal element. The action on fields give the familiar Buscher rules,
and  in double field theory this corresponds to the interchange
$x^a\leftrightarrow\tilde{x}_a$. 
In this formalism, one can still apply these transformations in the
case the generalized metric depends on the coordinate $x^a$, and the
resulting expression will still be a solution of the double field
theory equations of motion. Furthermore, the solution will be
compatible with the strong constraint provided one chooses
$(x^1,...,\tilde{x}_a,...,x^d)$ to be the physical coordinates.

One can now ask the question, to what extent these transformations give rise to equivalent string
theory backgrounds? The fact that such dualization of massive modes
could correct supergravity Buscher rules was emphasized and discussed
in \cite{Hull:2009mi} by using closed-string field theory. It should
be emphasized that such possibility, if true, is closely tied to the particular
toroidal background. We will discuss this possibility in connection of
the fibered structure of the semi-flat metrics and their quantum
corrections.


\subsubsection*{Dual backgrounds}

In the semi-flat limit, one can easily construct solutions of the
equations of motion derived from the action \eqref{eq:DFT-action} that
roughly correspond to the semi-flat limit of the doubled torus fibrations discussed in the
previous section. The semi-flat NS5-brane solution is lifted
to\,\footnote{We use $ds_{\text{DFT}}^2$ as a short-hand notation to encode the form of the generalized metric as $ds_{\text{DFT}}^2 = \mathcal H_{MN} dX^M dX^N$.}
\eq{
\label{eq:NS5semiflat-generalised-metric}
ds_{\text{DFT}}^2=\hspace{14pt}& h(r)\Bigl[ \op dr^2+r^2d\theta^2+(d\xi^1)^2+(d\xi^2)^2\Bigr]
\\
+\op&\frac{1}{h(r)}\left[\left(d\tilde{\xi}_1-\frac{\theta}{2\pi R_1 R_2}d\xi^2\right)^2+\left(d\tilde{\xi}_2+\frac{\theta}{2\pi R_1 R_2}d\xi^1\right)^2\right] .
}
This has a similar structure to the doubled torus fibration
\eqref{eq:doubletorusfibration} with
\begin{equation}
\tau= \frac{i}{2\pi\op R_1\op R_2}\log(z^{-1}) \, , \hspace{50pt} \tilde \tau = \frac{2\pi i \op R_1 R_2
}{\log(z^{-1})} \, ,
\end{equation}
where $z = r\op e^{i\theta}$, 
giving the expected monodromy $A_{\rho}$ \eqref{eq:doubledmonodromy}. 
By the simple basis change discussed above, one recovers the different semi-flat
backgrounds discussed in the previous section.

An interesting question is to what extent we can construct solutions
that incorporate the higher Fourier modes that localize the NS5-brane
on the torus fiber. In fact, this is possible by applying the generalized dualization
discussed in the previous subsection. This essentially corresponds to the
particular T-duality transformation \eqref{Erule} on the massive
fields. Starting from the NS5 solution on
$\mathbb{R}^2\times \mathbb T^2$, we obtain the following configuration
\eq{
\arraycolsep2pt
\label{eq:loc-KKmonopole}
\begin{array}{lcl}
ds^2&=&\displaystyle \tilde{h}\,\Bigl[dr^2+r^2d\theta^2+(d\xi^1)^2\Bigr]+\frac{1}{\tilde{h}}\left[d\xi^2+\frac{\tilde{R}_2}{2\pi R_1}\op \theta \,d\xi^1+\tilde{\Pi}_2\,d\theta\right]^2\,,\\[10pt]
B&=& \displaystyle \tilde{\Pi}_1\,d\theta\wedge d\xi^1 \, , \\[10pt]
e^{2\Phi} &= & {\rm const.}\,,
\end{array}
}
where $\tilde{\Pi}_{1,2}\equiv\Pi_{1,2}(r,\xi^1,\tilde{\xi_2})$ and $\Pi_{1,2}$ the functions collected in equation~\eqref{eq:B-field-2loc} in appendix~\ref{app:B-field}. Furthermore, $\tilde{h}\equiv h(r,\xi^1,\tilde{\xi}_2)$, with $h$ the localized harmonic function \eqref{eq:h-2loc}, and we write the result in terms of $\tilde{R}_2=1/R_2$.
This is a compactification of the solution presented in
\cite{Harvey:2005ab} and it is compatible with the naive T-duality
discussed in the previous sections.
A second dualization leads to the background described by
\eq{
\arraycolsep2pt
\label{eq:loc-Qbrane}
\begin{array}{lcl}
ds^2&=&\displaystyle \tilde{h}\op \bigl[ dr^2+r^2d\theta^2 \bigr]+\frac{4\pi^2\op\tilde{h}}{4\pi^2\tilde{h}^2+\tilde{R}_1^2\op\tilde{R}_2^2\op\theta^2}\left[(d\xi^1+\tilde{\Pi}_{1}\op d\theta)^2+(d\xi^2+\tilde{\Pi}_{2}\op d\theta)^2\right],
\\[10pt]
B&=& \displaystyle -\frac{2\pi\tilde{R}_1\tilde{R}_2\op\theta}{4\pi^2\tilde{h}^2+\tilde{R}_1^2\op \tilde{R}_2^2\op\theta^2}\op \bigl(d\xi^1+\tilde{\Pi}_{1}d\theta\bigr)\wedge\bigl(d\xi^2+\tilde{\Pi}_{2}d\theta\bigr) \,, 
\\[10pt]
e^{2\Phi} &=&\displaystyle \frac{4\pi^2\tilde{h}}{4\pi^2\tilde{h}^2+\tilde{R}_1^2\op\tilde{R}_2^2\op \theta^2} \,,
\end{array}
}
where now $\tilde{\Pi}_{1,2}\equiv\Pi_{1,2}(r,\tilde{\xi}_1,\tilde{\xi}_2)$ and $\tilde{h} \equiv h(r,\tilde{\xi}_1,\tilde{\xi}_2)$. We also substitute $R_{1,2}\rightarrow\tilde{R}_{1,2}=1/R_{1,2}$.
Configurations \eqref{eq:loc-KKmonopole} and \eqref{eq:loc-Qbrane}
depend explicitly on the winding coordinates, and should be understood
as DFT configurations by inserting the fields into
\eqref{eq:gen-metric} to obtain the corresponding generalized
metric. In the DFT language the above localized configurations can be
obtained from the localized NS5-brane generalized metric\,
by simple transformations of the type $\xi^a\leftrightarrow\tilde{\xi}_a$. All these configurations have 
vanishing generalized curvature $\mathcal{R}_{MN}$  and therefore are solutions of the equations of motion of the DFT action \eqref{eq:DFT-action}.

Before closing this section, let us mention that in principle one can
consider defects with more general $\tau$ and $\rho$ monodromies, of
the form of $(ADE, ADE)$ type. The physics of such non-geometric
defects has been recently studied in \cite{Font:2016odl}. Analogous
puzzles related to generalized T-duality will arise. However, in this
situation both winding and momentum might not be conserved along both
fiber directions, and a solution of the strong constraint is a priori not guaranteed.


\section{Discussion}
\label{sec_concl}

In this note we argued that winding modes are crucial for understanding the
near-core physics of T-duality defects. This is already evident in the
T-duality between $I_1$ singularities and NS5-branes, where our analysis becomes
essentially a compactified version of \cite{Gregory:1997te}. We argued
that a similar physics describes non-geometric defects, where an
essential role is played by two dyonic
coordinates dual to the isometric directions of the fiber torus.
In particular, the picture that emerges is that winding modes correspond to the 
localization of the Q-monopole compactified on a two-torus similarly as winding modes 
localize the KK-monopole compactified on a circle. 
It would be important to search for an explicit CFT description of
such winding mode physics, or a dual formulation in terms of more
conventional dynamics.

In the case of duality between the $I_1$
singularity and NS5-branes, there is an interesting observation
\cite{Israel:2004ir} (see also \cite{Martinec:2017ztd}). Consider a stack of $N$ NS5-branes on the
Coulomb branch, symmetrically distributed on a contractible
circle. There exists a particular large $N$ limit of such configuration in which
it looks like a periodic array of NS5-branes on a line. In such limit
the harmonic function is precisely of the form
\eqref{eq:Hmonopole}. Namely, the solution reduces to the $H$-monopole. 
Since we have an explicit CFT description of such a  NS5-ring, one
could understand the fate of the modes that localize the NS5-branes
after a T-duality. Unfortunately, to study T-duality to a
non-geometric configuration we would need to start with a
configuration of NS5-branes arranged on $S^1\times S^1$, for which an
explicit CFT description is lost.

A different approach to study torus fibrations is to 
fiber over a common base $\mathcal{B}$ the eight dimensional duality between the
heterotic string compactified on a torus and F-theory compactified on
a elliptic K3-surface. If we keep the gauge group unbroken the eight-dimensional 
moduli are $\tau$ and $\sigma$, and a fibration over
$\mathcal{B} = \mathbb{P}^1$ reproduces our models. The $\tau$- and
$\sigma$-fibrations are described by two Weierstrass forms  
\begin{equation}
\tau : \, y^2 = x^3 + f_{\tau} \op x + g_{\tau} \, ,\hspace{50pt}
\rho: \,  \tilde y^2 + \tilde x^3 +
f_{\rho}\op \tilde x + g_{\rho} \, ,
\end{equation}
from which we can reconstruct the dual elliptic K3. Note that the NS5-brane, 
as well as the Q-brane and the general $(p,q)_{\rho}$-branes, are
identified with a $I_1$ degeneration of the $\rho$-fibration. In fact,
the explicit map to F-theory is completely symmetric between $\tau$ and
$\rho$, as expected from T-duality. This however leads to a
puzzle if we consider the leading symmetries of the exact solutions,
namely the corrections to the semi-flat approximation. The $I_1$
degeneration of the $\tau$-fibration correctly captures the breaking
of the $U(1)$ isometry of the cycle which is orthogonal to the
shrinking one. This is described by the metric
\eqref{eq:oogurivafa}. On the contrary, the
$I_1$ degeneration of the auxiliary $\rho$-fibration misses the
symmetries of the NS5-solution. Additionally, the $(p,q)_{\rho}$
conjugacy class interpolates between the NS5-brane, where both the
$U(1)^2$ isometries of the torus are broken, and the Q-brane, where
they are expected to be exact. How these facts are reconciled with T-duality?
In fact, the moduli corresponding to
the position of the NS5-branes on the fiber are missed by the
heterotic/F-theory duality. The explicit solutions on the doubled
space that we obtained in section~\ref{sec_dft} (assuming they are not
substantially modified for the heterotic string) give a precise
prediction for the corrections to the semi-flat $\rho$-fibration. 
Hence, it should be possible to check whether they capture
the essential physics by extending the duality to F-theory to include
these extra moduli. 
It would also be interesting to see if some of these results can be
obtained by using D-brane probes, generalizing the analysis of \cite{Witten:2009xu}.


\vspace{0.5cm}
\subsubsection*{Acknowledgments} 

We are grateful to Stefano Massai for collaboration on this project.
We furthermore thank Ismail Achmed-Zade, Daniel Junghans and Felix Rudolph for valuable discussions. 
This work was partially supported by the ERC Advanced Grant ``Strings and Gravity''
(Grant.~No. 320045) and by the DFG cluster of excellence ``Origin and Structure of the Universe''.  


\clearpage
\appendix

\section{Domain walls}

In this appendix, we briefly consider the case in which the space transversal
to the monopoles is $\mathbb{R}\times \mathbb T^3$. Let us begin with
the NS5-brane, for which the exact metric can be determined as before by
considering a three-dimensional array of harmonic sources
\begin{equation}
h(y,\xi^1,\xi^2,\xi^3)=\sum_{a=1}^3\sum_{\vec{n}\in\mathbb{Z}^3}\frac{1}{y^2+(\xi^a-2\pi R_a n_a)^2}\,,
\end{equation} 
where $y$ is the coordinate on $\mathbb{R}$ and $\xi^i$ are
coordinates on the $\mathbb T^3$ with periodicities $\xi^i\sim \xi^i+2\pi R_i$. The sum does not converge, but can be regularized. After a Poisson resummation it becomes
\begin{equation}
h(y,\xi^1,\xi^2,\xi^3)=\frac{1}{4\pi R_1R_2R_3}
\left[|y|+\sum_{a=1}^3\sum_{\vec{k}\in(\mathbb{Z}^3)^{^*}}\frac{1}{\lambda}e^{-\left(i\frac{k_a\xi^a}{R_a}+\lambda |y|\right)}\right],
\end{equation}
where 
\begin{equation}
\lambda=\sqrt{\left(\frac{k_1}{R_1}\right)^2+\left(\frac{k_2}{R_2}\right)^2+\left(\frac{k_3}{R_3}\right)^2}.
\end{equation} 
In the limit $|y|\gg R_1,R_2,R_3$ the branes are smeared on the torus
and the solution is simply
\eq{
 \label{eq:NS5onT3}
\arraycolsep2pt
\begin{array}{lcl}
ds^2 &=&\displaystyle  h(y) \Bigl[ dy^2 + (d\xi^1)^2+ (d\xi^2)^2+ (d\xi^3)^2\Bigr]   \, ,\\[6pt] 
H &= &\displaystyle -\frac{\text{sgn}(y)}{4\pi R_1R_2R_3}\, d\xi^1\wedge
    d\xi^2\wedge d\xi^3 \,, \\[10pt]
e^{2\Phi} &=&  h(y) \, ,
\end{array}
}
with $ h(y)= |y|/(4\pi R_1R_2R_3)$. A T-duality along say the
$\xi^2$-direction gives a background of the form $\mathbb{R}\times
{\rm Nil}_3$, which is in fact a Taub-NUT space \eqref{eq:ghansatz} with
special circle $\xi^2$, where two of the
base directions have been compactified. The smeared limit is given by
the metric
\begin{equation}\label{eq:Nilwall}
ds^2 = h(y)\Bigl[ dy^2 + (d\xi^1)^2+(d\xi^3)^2\Bigr] +
\frac{1}{h(y)}\op \Bigl[ d\xi^2-\frac{\text{sgn}(y)}{4\pi R_1 R_2 R_3}\xi^3 d\xi^1\Bigr]^2 \, .
\end{equation}
At fixed $y$, there is a torus parametrized
by $(\xi^1,\xi^2)$ which is fibered over the cycle corresponding to $\xi^3$, with
monodromy given by a Dehn twist around $\xi^1$ resulting in a nilmanifold as a total
space. For a detailed discussion of the domain-wall metric \eqref{eq:Nilwall}
see for example \cite{Gibbons:1998ie,Lust:2015yia}.
 As
for the elliptic cases discussed in the main text, an additional T-duality
along say the $\xi^1$-direction gives a solution in which, at fixed
$y$, there is a non-geometric T-fold given by the $(\xi^1,\xi^2)$-torus fibration over $\xi^3$ 
with monodromy $1/\rho \rightarrow
1/\rho -1$. 
This T-fold solution has no additional isometry, but it has been speculated
from considerations in the effective reduced theory that a last T-duality is possible,
giving a so-called R-space \cite{Shelton:2005cf,Hassler:2013wsa}. From the present
perspective it seems that their physics in the closed-string sector would be 
essentially determined by the T-dual of
the massive modes in \eqref{eq:NS5onT3}.


\section{$B$-fields of the localized NS5-brane}
\label{app:B-field}

For completeness, in this appendix we give  the full expressions of the $B$-field for the localized
NS5-brane solutions.


\subsubsection*{NS5 on $\mathbb{R}^3\times S^1$}

Consider an NS5-brane on $\mathbb R^3\times S^1$. 
Using spherical coordinates for the non-compact transversal space of the form
\begin{equation}
\bigl(\op\vec{x} \op,\op\xi^2\op\bigr)=\bigl(\op |\vec{x}| \sin\theta\cos\varphi \op,\op|\vec{x}|
\sin\theta\sin\varphi\op,\op|\vec{x}| \cos\theta\op,\op\xi^2\op \bigr),
\end{equation}
the $B$-field up to gauge transformations reads
\begin{equation}
B=\frac{1-\cos\theta}{2 R_2}\,d\varphi\wedge d\xi^2+B_{\theta\varphi }\,d\theta \wedge d\varphi \,,
\end{equation}
with
\eq{
\label{eq:B-field-4loc}
B_{\theta\varphi}=\sin \theta \left[\tan ^{-1}\left(\coth \frac{|\vec{x}|}{2 R_2} \tan \frac{\xi^2}{2 R_2}\right)-\frac{\xi^2}{2 R_2}+\frac{|\vec{x}| \sin \left(\xi^2/R_2\right)}{2 R_2 \cosh \left(|\vec{x}|/R_2\right)-2 R_2 \cos \left(\xi^2/R_2\right)}\right] .
}
In the semi-flat limit, $B_{\theta\varphi}\rightarrow 0$ and the B-field simplifies to
\begin{equation}\label{eq:B-field-3}
B^{\text{sf}}=\frac{1-\cos\theta}{2\op R_2}\, \,d\varphi\wedge d\xi^2.
\end{equation}


\subsubsection*{NS5 on $\mathbb{R}^2\times \mathbb T^2$}

For an NS5-brane on $\mathbb R^2\times\mathbb T^2$ we have the following expression for the 
$B$-field
\begin{equation}
B=\frac{\theta}{2\pi R_1 R_2}\,d\xi^1\wedge d\xi^2+\Pi_{1}\,d\theta\wedge d\xi^1+\Pi_{2}\,d\theta\wedge d\xi^2 \, \,,
\end{equation}
where $\Pi_{1,2}$ satisfy the equations
\eq{ \partial_r\Pi_1=r\partial_2h\, ,
\hspace{40pt} 
\partial_r\Pi_2=-r\partial_1h\, ,
\hspace{40pt}
(\partial_2\Pi_2-\partial_1\Pi_1)= r\partial_rh+(2\pi R_1 R_2)^{-1}\, ,
}
where $h$ is the localized harmonic function \eqref{eq:h-2loc}. These equations are solved by
\begin{align}
\label{eq:B-field-2loc}
\Pi_{1}(r,\xi^1,\xi^2)&=+\sum_{k_1,k_2\geq0}\frac{(2-\delta_{k_1,0}-\delta_{k_2,0})}{\pi R_1 R_2^2}\, \frac{k_2}{\lambda}\, r\, K_1(\lambda r)\cos \left(\frac{k_1\xi^1}{R_1}\right) \sin \left(\frac{k_2\xi^2}{R_2}\right)  , \\
\Pi_{2}(r,\xi^1,\xi^2)&=-\sum_{k_1,k_2\geq0}\frac{(2-\delta_{k_1,0}-\delta_{k_2,0})}{\pi R_1^2 R_2}\, \frac{k_1}{\lambda}\, r \, K_1(\lambda r)\sin \left(\frac{k_1\xi^1}{R_1}\right) \cos \left(\frac{k_2\xi^2}{R_2}\right) , 
\end{align}
with $K_1$ the first order modified Bessel function of second kind and
$\lambda=\sqrt{\left(k_1/R_1\right)^2+\left(k_2/R_2\right)^2}$.


\vspace{0.5cm}
\bibliographystyle{utphys}
\bibliography{bibliounwinding}


\end{document}